\AtBeginDocument{%
  }

\documentclass[sigconf]{acmart}

\usepackage{amsmath, amssymb, algorithm, algpseudocode,multirow}
\copyrightyear{2025}
\acmYear{2025}
\setcopyright{acmlicensed}
\acmConference[KDD '25] {Proceedings of the 31st ACM SIGKDD Conference on Knowledge Discovery and Data Mining V.2}{August 3--7, 2025}{Toronto, ON, Canada.}
\acmBooktitle{Proceedings of the 31st ACM SIGKDD Conference on Knowledge Discovery and Data Mining V.2 (KDD '25), August 3--7, 2025, Toronto, ON, Canada}
\acmISBN{979-8-4007-1454-2/25/08}
\acmDOI{10.1145/3711896.3737195}

\begin{document}


\title{Audio-Enhanced Vision-Language Modeling with Latent Space Broadening for High Quality Data Expansion}

\author{Yu Sun}
\affiliation{%
  \institution{ByteDance Inc.}
  \city{San Jose}
  \state{CA}
  \country{United States}
}
\email{yu.sun@bytedance.com}

\author{Yin Li}
\affiliation{%
  \institution{ByteDance Inc.}
  \city{San Jose}
  \state{CA}
  \country{United States}
  }
\email{liyin.rd@bytedance.com}

\author{Ruixiao Sun}
\affiliation{%
  \institution{ByteDance Inc.}
  \city{San Jose}
  \state{CA}
  \country{United States}
  }
\email{ruixiao.sun@bytedance.com}

\author{Chunhui Liu}
\affiliation{%
  \institution{ByteDance Inc.}
  \city{San Jose}
  \state{CA}
  \country{United States}
  }
\email{chunhui.liu@bytedance.com}

\author{Fangming Zhou}
\affiliation{%
  \institution{ByteDance Inc.}
  \city{Beijing}
  \state{Beijing}
  \country{China}
  }
\email{zhoufangming@bytedance.com}

\author{Ze Jin}
\affiliation{%
  \institution{ByteDance Inc.}
  \city{Bellevue}
  \state{WA}
  \country{United States}
  }
\email{ze.jin@bytedance.com}

\author{Linjie Wang}
\affiliation{%
  \institution{ByteDance Inc.}
  \city{San Jose}
  \state{CA}
  \country{United States}
  }
\email{wanglinjie.w@bytedance.com}

\author{Xiang Shen}
\affiliation{%
  \institution{ByteDance Inc.}
  \city{Bellevue}
  \state{WA}
  \country{United States}
  }
\email{xiang.shen@bytedance.com}

\author{Zhuolin Hao}
\affiliation{%
  \institution{ByteDance Inc.}
  \city{Beijing}
  \state{Beijing}
  \country{China}
  }
\email{haozhuolin@bytedance.com}

\author{Hongyu Xiong}
\authornote{Author corresponded for this research.}
\affiliation{%
  \institution{ByteDance Inc.}
  \city{San Jose}
  \state{CA}
  \country{United States}
}
\email{hongyu.xiong@bytedance.com}

\renewcommand{\shortauthors}{Yu Sun et al.}

\begin{abstract}

Transformer-based multimodal models are widely used in industrial-scale recommendation, search, and advertising systems for content understanding and relevance ranking. Enhancing labeled training data quality and cross-modal fusion significantly improves model performance, influencing key metrics such as quality view rates and ad revenue.
High-quality annotations are crucial for advancing content modeling, yet traditional statistical-based active learning (AL) methods face limitations: they struggle to detect overconfident misclassifications and are less effective in distinguishing semantically similar items in deep neural networks. Additionally, audio information plays an increasing role, especially in short-video platforms, yet most pretrained multimodal architectures primarily focus on text and images. While training from scratch across all three modalities is possible, it sacrifices the benefits of leveraging existing pretrained visual-language (VL) and audio models.
To address these challenges, we propose kNN-based Latent Space Broadening (LSB) to enhance AL efficiency, achieving an up to 9\% recall improvement at 80\% precision on proprietary datasets. Additionally, we introduce Vision-Language Modeling with Audio Enhancement (VLMAE), a mid-fusion approach integrating audio into VL models, yielding up to another 9\% recall improvement at 80\% precision. Our methods are successfully deployed in multiple production systems, leading to significant business gains through online A/B experiments.

\end{abstract}





\maketitle

\section{Introduction}

In modern search, recommendation, and advertising systems, multimodal content understanding models play a critical role in evaluating content quality and determining the relevance between search queries and candidate items~\cite{devlin2018bert, hou2024crossdomainlifelongsequential,sharma2024optimizingnoveltytopkrecommendations, wang2025reasoningenhanceddomainadaptivepretrainingmultimodal}. For example, feed recommendation heavily relies on content understanding models to suppress content that can lead to negative user experiences and promote content aligned with users' interests. In addition, search advertising systems depend on relevance models to assess the degree of relevance between a given query-ad pair. To enable multimodal information integration across visual, textual, and audio data, multimodal models have become widely adopted, significantly enhancing content understanding capabilities. Improving data quality and modality fusion are two critical directions for further advancing multimodal model performance.




Training data for these models are heavily based on manually annotated samples. However, random sampling-based annotation pipelines often result in inefficiencies, particularly for tasks where positive samples are sparse while labeling costs are high. This imbalance can lead to an excessive proportion of easy instances that the model has already correctly classified, limiting the effectiveness of additional annotations.

Active Learning (AL) aims to improve annotation efficiency by prioritizing difficult samples for labeling, making model training more effective. Conventional statistical-based AL strategies select samples based on multiclass output, such as Least Confident ~\citep{culotta2005reducing}, Margin Sampling~\citep{scheffer2001active}, and Max Entropy ~\citep{kim2006mmr}. These statistical methods primarily focus on uncertainty sampling~\citep{settles2008analysis}, prioritizing samples where the model exhibits low confidence, as indicated by close probability scores among classes or high entropy in the output distribution.

However, statistical AL strategies alone are insufficient for deep neural networks (DNNs). First, the final multiclass probability vector is derived from high-dimensional hidden embeddings, where the compression of semantic information may obscure fine-grained distinctions, making it difficult to detect semantically similar yet hard-to-classify samples. Furthermore, due to the complex interplay between model representations and data uncertainty, it is challenging to identify instances where the model is confident but incorrect. Thus, expanding AL strategies to latent space representations is crucial for further improving annotation efficiency.

Furthermore, modality fusion presents significant challenges for short-video applications where video, text, audio modalities often contain non-overlapping information and audio features plays an critical role in enhancing content understanding models. The success of multimodal models often benefits from large-scale pretraining. However, the video-text~\cite{zhao2024multimodal}, video-audio~\cite{liu2023cat,zhu2024meerkat} and text-audio~\cite{radford2022whisper} models are more prevalent while pretrained models~\cite{liu2023macaw} with all three modalities based on social media data remain scarce. On the other hand, traditional methods for audio content extraction, such as Automatic Speech Recognition (ASR), is also not feasible due to two key challenges: 1) dependency on content extraction model quality, often struggle with multilingual processing and may extract text unrelated to video content; 2) converting ASR into text significantly increases the input token sequence length for vision-language models, adding to the model complexity, and truncation may lead to information loss. Thus, the development of efficient mid-to-late-fusion strategies that enable direct interaction between audio, video, and text modalities presents a promising direction for improving content understanding models.

To address these two challenges, this paper presents two key contributions:

(1) A novel active learning strategy based on \textbf{Latent Space Broadening (LSB)} – By identifying semantically similar hard samples from a set of seed bad cases, we leverage k-Nearest Neighbors (kNN) in the latent space, which is derived from the hidden state embeddings of the multi-modality models. We further introduce a Lookalike Threshold (LT) to quantitatively define embedding similarity, enhancing active learning effectiveness\cite{chacko2021customer, zhang2025beyond}.

(2) \textbf{Vision-Language Modeling with Audio Enhancement (VLMAE)} – We compare a genuine approach of audio fusion~\cite{dong2024coefvq, zhao2024balf} and propose VLMAE, which facilitates the modality crossing between audio and vision-language (VL) information, then further perform supervised fine-tuning incorporating domain-specific knowledge. This approach significantly improves the model’s ability to jointly understand video, text, and audio.

We conduct extensive experiments on 3 different tasks from content quality in the recommendation system and search ads relevance, and demonstrate promising performance improvements across offline metrics, case analyses, and online A/B testing results. We have deployed the proposed approaches in multiple production systems after A/B experiments.





\section{Related Work}

\textbf{Active Learning}.
The theoretical foundations of active learning algorithms have been extensively studied~\citep{dasgupta2005analysis, balcan2009agnostic, awasthi2014power, yan2017revisiting, beygelzimer2009importance, zhou2024information}. 
However, statistical algorithms guarantees are often not directly applicable to deep neural networks (DNNs). 
Due to these limitations, practical active learning applications often rely on heuristic approaches to select examples for labeling. For instance, \citet{tong2001support} and \citet{SUN2023106742} propose a margin-based selection strategy, while \citet{settles2008analysis} and \citet{shen2004multi} suggest combining multiple criteria for natural language processing (NLP) tasks. \citet{culotta2005reducing} explore the use of the least confidence criterion in linear CRF models for sequence prediction tasks. 
A broader overview of the field can be found in the works of \citet{settles2010active} and \citet{olsson2009literature}.

Active learning has been extensively studied in content understanding model. ~\citet{gal2017deep} applied Bayesian neural networks to enhance sample selection for high-dimensional data, while ~\citet{sinha2019variational} proposed a variational adversarial framework to learn latent representations for effective sampling. In NLP, ~\citet{cossu2017active} leveraged active learning to annotate micro-blogs for e-reputation tasks, tackling domain-specific challenges. ~\citet{ash2020deep} introduced BADGE, which selects diverse and uncertain examples using gradient embeddings and K-means, bridging uncertainty, and representation diversity. These works highlight the importance of embedding spaces and uncertainty modeling in improving active learning efficiency. \citet{xie2021fast}\citet{yoo2019learning} also explored the Active learning application on loss prediction module to estimate the informativeness of unlabeled samples.




\textbf{Lookalike Threshold (LT) Modeling}.
The study by \citet{chacko2021customer} explores various machine learning lookalike techniques aimed at identifying similar customer profiles, providing insights into how data-driven models can enhance targeted marketing strategies.
Collectively, these studies underscore the potential of integrating kNN and lookalike methodologies to enhance the performance and adaptability of NLP models across various tasks and platforms.

\textbf{Visual-language Models}.
The pretraining of transformers for various visual-language processing tasks has seen substantial improvements in recent years such as X-VLM~\cite{wang2022multi}, VLP~\cite{vlp2020}, SimVLM~\cite{simvlm2021}. 
The BEiT model~\cite{bao2022beitbertpretrainingimage} introduced a vision transformer pretraining approach that focuses on leveraging large-scale unlabeled data and improving the model's ability to handle downstream tasks such as image classification. 
Building upon this, the VLMo~\cite{li2021vlmo} model takes the concept of vision transformers a step further by integrating more complex modalities with Mix of Experts (MoE) framework. 
In the context of fine-tuning for specific tasks, it has been shown that multimodal models can be adapted effectively for both language and vision-related applications, thereby providing state-of-the-art results across multiple benchmark datasets.

\textbf{Audio Modality}.
In the area of speech recognition, OpenAI's Whisper model \cite{radford2022whisper} has demonstrated remarkable performance by utilizing large-scale weak supervision for speech-to-text tasks. 
The model is trained on a wide variety of languages and audio formats, allowing it to generalize well to diverse speech recognition challenges. 
Whisper's framework aligns with the recent trend of using massive datasets to drive the performance of deep learning systems in audio processing.

\textbf{Sequence-Based Cross Network}.
Another important area in recent research involves recommendation systems, where models such as DIN(Deep Interest Network) \cite{liu2022pretrain}, DIEN\cite{dien2019}, DIHAN\cite{dihan2021}, have gained attention for their ability to effectively model user interests and behavior, even the sequence information from more general inputs, such as vectors. The sequence-based framework utilizes deep learning techniques to improve recommendation accuracy by understanding the user's historical interactions with items or other sequential information. 
Moreover, its flexibility in handling dynamic user behavior not only makes it a popular choice for real-time recommendation tasks, the interaction of multi-dimensional vectors can also be applied to the interaction of all serialized information or embeddings. This advancement marks a shift towards utilizing deeper, more context-aware architectures for improving recommendation systems.

\section{Approaches}
\label{sec:search_modeling}
A generic vision-language (VL) modeling for video classification nowadays employs a workflow like this: 
(1) First, takes video frames as input and projects to visual tokens under the same representation space as text tokens;
(2) then fuses both visual and textual tokens with a transformer-based architecture;
(3) finally uses the embeddings generated from the transformer for classification prediction. ~\citep{li2021vlmo, wang2022image}


\textbf{Visual Token Projection.}
For the sequence of sampled video frames with length $T$, the raw input $X_f$ is passed through a vision layer $g(\cdot; \theta)$ which extracts its visual features and projects to a modality-aligned representation space: 
\begin{equation}
    H_f^{(i)} = g(X_f^{(i)}; \theta)
\end{equation}
, where $i = 1, 2, ..., T$, and $H_f$ serves as the sequence of visual tokens, facilitating downstream integration with textual tokens.

\textbf{Fusion of Visual and Text Tokens.}
Then a VL fused layer processes a sequence of both visual tokens \( H_f \) and text tokens \( X_t \) (e.g. query, and the video's textual information). 
Typically, the VL-fused transformer $VLT(\cdot; \theta)$ takes the multi-modal tokens as input and generates a sequence of hidden representations, $H$, where each representation corresponds to an input token, maintaining the same sequence length as the input sequence. 
\begin{equation}
    H = VLT([H_f;X_t]; \theta)
\end{equation}

\textbf{Classification Module}
The classification module $f_\mathrm{cl}(\cdot; \theta_\mathrm{cl})$ is applied on top of the hidden state embedding \(H_{cls}\) of the [CLS] token. 
This module consists of a Multilayer Perceptron (MLP) mapping $H_{cls} \in {R}^{d}$ to ${R}^{|\mathcal{Y}|}$, where $\mathcal{Y}$ is the set of target classes. 

\subsection{Latent Space Broadening}
Content understanding models highly rely on human-annotated training data. 
However, the volume of human annotations is inherently limited, allowing only a sampled subset of data on the internet platform to undergo annotation. 
It is imperative to utilize the limit annotation resource smartly by selecting the samples which are most effective for improving the model performance to annotate.

\begin{figure*}[h] 
  \centering
  \includegraphics[width=0.8\textwidth]{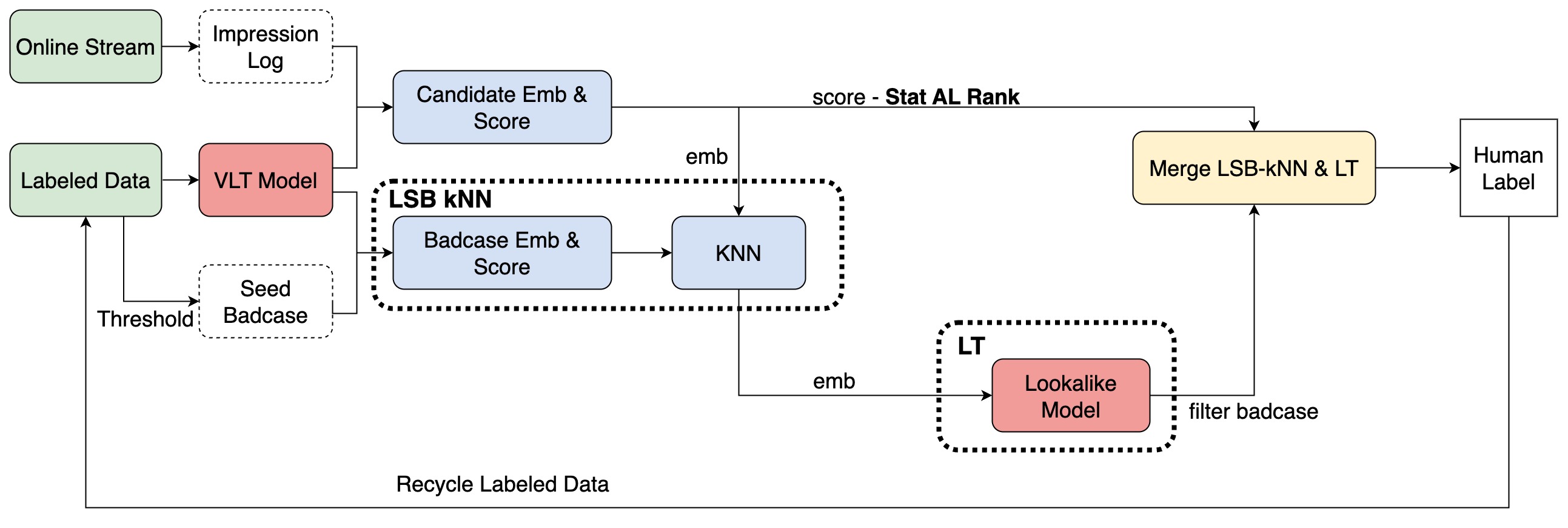} 
  \caption{Active Learning Pipeline with kNN-based LSB under LT. 
  In this pipeline, candidate samples are randomly selected from impression logs, and a set of seed badcases are selected by comparing the human annotation and the vision-language transformer (VLT) model prediction. 
  Based on kNN, seed badcases are expanded by choosing samples from candidate set which have the closest hidden embeddings.
  The broadened set is further filtered through a binary-class lookalike model given a LT value. 
  The resulting badcases are then mixed with statistical-AL-selected candidates, forming the dataset for annotation for next round of VLT model updates.}
  \label{fig:flowchart}
\end{figure*}

As illustrated in Figure \ref{fig:flowchart}, we have designed an active learning (AL) pipeline to manage annotations recurrently, continually refining AL sampling strategies based on the latest snapshot of the model. (Algorithm \ref{algo})

\textit{Seed Bad Cases Preparation.} We define "bad-case" as situations where the model's prediction is significantly misaligned with human annotations.
From the most recent date partitions of the impression log (usually the last 30 days), we sort the bad cases based on the difference between the annotated label and the predicted score of the model, 
and then cache the CLS hidden embeddings and the final multi-class vectors of the corresponding samples for further analysis and improvement. 
The top bad cases with the largest discrepancy are selected as "Seed Bad Cases".

\textit{Bad Cases Expansion through latent space broadening (LSB).}
Based on the k-Nearest-Neighbor (kNN) algorithm, we select similar cases whose hidden embeddings \(H_{cls}\) are similar to those of seed cases. 



›\begin{algorithm}[h]
\caption{\textbf{LSB-LT:} Active Learning Annotation through Latent Space Broadening under Lookalike Threshold}
\label{alg:lsb-lt}
\begin{algorithmic}[1]
\Require
\begin{itemize}
    \item VLT model $\mathcal{M}(x; \theta)$
    \item Initial annotated dataset $D_0$
    \item Unlabeled candidate pool $U_i$ (can be recurrently collected)
    \item Lookalike model $\mathcal{G}(h_\theta; \phi)$
    \item Threshold $s_t$ (for seed badcase selection)
    \item Threshold $l_t$ (for lookalike filtering)
    \item Maximum number of rounds $T$, new round is triggered given recurrence requirement.
\end{itemize}
\For{$i = 0$ to $T$}
  \State \textbf{Train VLT model:}
    Train and update $\mathcal{M}$ to obtain parameters $\theta_i$, using dataset $D_i$.
    
  \State \textbf{Find seed badcases:}
    \[
      S_{i+1} \gets \left\{ x \in D_i \;\middle|\; \bigl|\mathcal{M}(x; \theta_i) - \mathrm{label}(x)\bigr| > s_t \right\}
    \]

  \State \textbf{Obtain LSB set via kNN:}
    Perform kNN on $U_i$ with respect to $S_{i+1}$ to obtain $S_{i+1}^{\mathrm{LSB}}$.

  \State \textbf{Update Lookalike Model:}
    Train/update $\mathcal{G}$ based on $S_{i+1}$, resulting in parameters $\phi_{i+1}$.

  \State \textbf{Lookalike threshold filtering:}
    \[
      S_{i+1}^{\mathrm{LSB-LT}} \gets \left\{ x \in S_{i+1}^{\mathrm{LSB}} \;\middle|\; \mathcal{G}\bigl(h_{\theta_i}(x); \phi_{i+1}\bigr) < l_t \right\}
    \]
    Annotate samples in $S_{i+1}^{\mathrm{LSB-LT}}$ and obtain incremental dataset:
    \[
      D_{i+1} \gets  \mathrm{Annotate}\bigl(S_{i+1}^{\mathrm{LSB-LT}}\bigr)
    \]
\EndFor
\end{algorithmic}
\label{algo}
\end{algorithm}

\subsubsection{kNN on Latent Embedding}
\textbf{kNN (k-Nearest Neighbors)} algorithm facilitates the identification of similar cases to the seed badcase by calculating similarity between embeddings in latent space. 
This enables the preservation of semantic similarity. 

Based on the final classification module (usually a two-layer MLP), we use $h$ to simplify $H_{cls}$ describled in Sec~\ref{sec:search_modeling}:
\begin{equation}
\hat{y} = f(h;W) = \text{softmax}(W \cdot h )
\label{softmax}
\end{equation}
; for any $h'$ belongs to kNN of $h$, we would expect that under a certain limit $\epsilon$, s.t. 
$\| h' - h \| \leq \epsilon$, there would be
$\| \hat{y}' - \hat{y} \| \leq \delta$, s.t. within $\delta$-vicinity of \( \hat{y} \), \( \hat{y}' \) would give the same output class of \( output = \arg\max_i \hat{y} \).

However, due to the non-linearity nature of $f(h;W)$, it is difficult for the kNN algorithm to explicitly guarantee that $h$ and $h'$ are close enough in the latent space.
In addition, even though for a certain $h'$ we are able to guarantee the $\delta$-vicinity, its actual label $l' = \text{idx}(y')$ could be different from the label $l = \text{idx}(y)$ of $h$, 
i.e. under search scenario, two pairs of query-videos could be semantically close, but one is "$prediction-label\; match$" but the other is "$prediction-label\; mismatch$". 
In this situation, we are not able to expand to as many bad cases as expected.

Therefore, in the next section we introduce a way to directly model the gap between annotation and the model prediction, and then use this quantitative measure as a threshold to enhance the quality of kNN.

\subsubsection{Lookalike Threshold (LT)}
\label{sec:lookalike}
To address the limitation of kNN discussed in the previous section, we propose Lookalike modeling~\cite{shen2015effective,chacko2021customer} to quantitatively rank the kNN-selected samples, 
and to enhance the overall quality of LSB for bad case expansion by applying a lookalike threshold (LT).

In this method, taking the latent embedding as input, a binary classification model is trained to distinguish whether there is a mismatch between the prediction of the model $output$ and the human annotation $l$. 

Here, we define label \( d = 1 \) as inconsistent between $output$ and $l$ ($\arg\max_i \hat{y} \neq \text{idx}(y)$), and \( d = 0 \) as consistent between $output$ and $l$ ($\arg\max_i \hat{y} = \text{idx}(y)$); a logistic regression (LR) model is trained with variable $U$ and $b$: 
\begin{equation}
    \hat{d} = g(h) = \sigma(U \cdot h + b)
\end{equation}
; for all possible \( h' \), 
s.t. $\hat{d}' = g(h') > \text{threshold}$, we could use the threshold to adjust the quality of LSB.
(Figure \ref{fig:flowchart}).

\subsection{Vision-Language Modeling with Audio Enhancement (VLMAE)}
We propose to incorporate audio modality by integrating the output of a pre-trained audio encoder into the VL architecture and effectively interacting with VL modalities through a sequence-based crossing network.

\textbf{Pre-trained Audio Encoder for Video’s Audio.} To incorporate audio information, a pre-trained audio encoder \( a(\cdot; \theta) \) processes the raw audio input from the video, denoted as \( X_a \). A sequence of hidden representation \( H_{A} \) of the processed audio signal is generated.
\begin{equation}
    H_{A} = a(X_a; \theta)
\label{encode}
\end{equation}
This audio encoder extracts audio features relevant to the classification task, allowing the model to incorporate sound cues, which are particularly useful when videos include narration or other informative audio elements.

During modality fusion, a genuine approach is to directly concatenate the average or max pooling of the audio embedding sequence with the final CLS embedding from the VL model (Eq~\ref{softmax}). 
\begin{equation}
    \hat{y} = \text{MLP}(\operatorname{concat}(H_{cls}, \text{avg\_pooling}(H_{A})))
\label{concat}
\end{equation}
This is a classic late-fusion approach~\cite{dong2024coefvq}, without sufficient interaction between audio and VL information.
Besides, this method suffers from the over-compression of audio information due to average/max pooling. 

\begin{figure}[h] 
  \centering
  \includegraphics[width=0.5\textwidth]{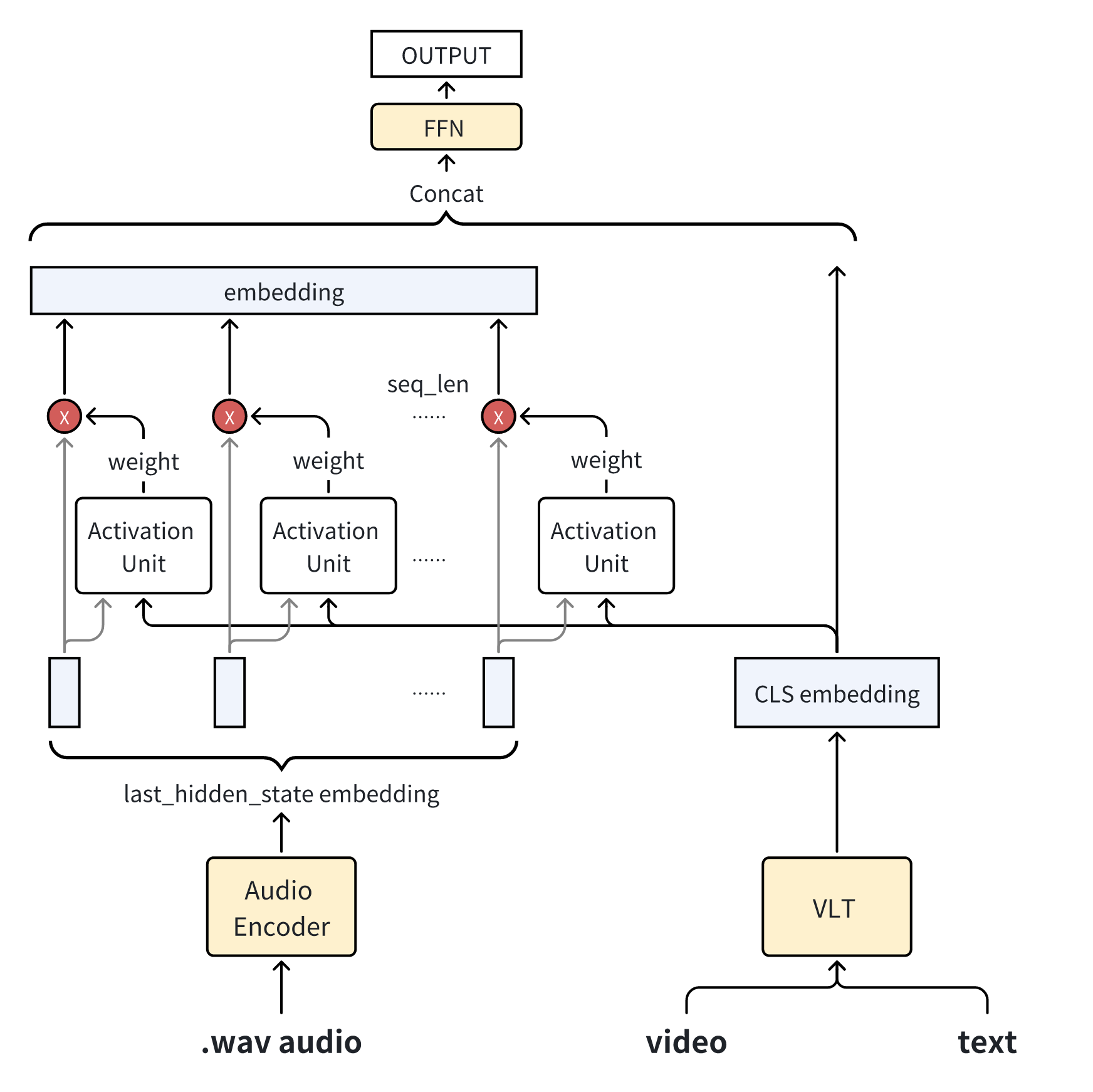} 
  \caption{Vision-Language Modeling with Audio Enhancement (VLMAE) where a learnable attention layer is introduced to improve the fusion between audio and VL information. This enables the model to effectively fit and recognize audio modality features.}
  \label{fig:DIN}
\end{figure}

\textbf{Vision-Language Modeling with Audio Enhancement (VLMAE)}
To address this problem, we propose the Vision-Language Modeling with Audio Enhancement (VLMAE) approach. 
By using the CLS embedding from the VL model as an anchor, we introduce a learnable attention layer upon the audio embedding sequence, 
and the attention for audio embedding within the sequence is calculated and normalized with respect to the VL CLS embedding. 
This allows a more sufficient crossing between audio and VL information before entering the MLP, thereby facilitating fusion between the modalities. (Figure\ref{fig:DIN})

\begin{equation}
\hat{y} = \text{MLP(}\text{concat}\left(\text{softmax}\left(\frac{H_{A} \cdot ({H_{cls}})^\top}{\sqrt{d_A}}\right) \cdot H_{A}, \, {H_{cls}}\right))
\label{tab:DIN}
\end{equation}
, where $d_A$ stands for attention dimension

\section{Experiments}
In this section, we conduct extensive offline and online experiments on three proprietary tasks to validate the effectiveness of the proposed approaches. 
The three tasks are Search Ads Relevance (Search Ads), Clickbait Video Detection (Clickbait), and Adult Nudity and Sexual Activities Video Detection(ANSA), respectively.
Both training and testing data in the offline experiments are manually annotated.
There is no data leakage between the evaluation set and the training set.

\subsection{Datasets}

\subsubsection{Evaluation Sets} \

\textbf{Search Ads}: 12.7k randomly sampled query-ad pairs from the past 30 days Search Ads impression log. Positive sample rate = $2.6\%$.

\textbf{Clickbait}: 40k positive concentrated sampled videos from the past 60 days daily published new video log. Positive sample rate = $12\%$.

\textbf{ANSA}: 30k positive concentrated sampled videos from the past 90 days daily published new video log. Positive sample rate = $26.7\%$.

\subsubsection{Training Sets}\

\textbf{Search Ads}: 
Choosing from the past 60 days query-ad impression logs, 
the dataset includes 1.72 million randomly sampled data, 330k statistical AL (96k least-confident, 56k margin-sampling, 181k max entropy) and 130k kNN-LSB.

For comparison, we pre-train and warm-up the model by 1.56 million random samples as a fixed snapshot, and the remaining 160k samples for each group are selected using different active learning (AL) methods to evaluate the effectiveness of various strategies.

\textit{Random}: the remaining 160k randomly sampled data serves as a baseline to demonstrate the effectiveness of AL in general.

\textit{Statistical AL}: 160k data sampled from 330k statistical AL 
based on 4-class output confidence distributions. 
These statistical methods primarily focus on uncertainty sampling~\citep{settles2008analysis}, prioritizing samples where the model exhibits low confidence, as indicated by close probability scores among classes or high entropy in the output distribution. (Eq~\ref{tab:leastconfident}, ~\ref{tab:marginsampling}, ~\ref{tab:entropy}), serves as another baseline to demonstrate the effectiveness of the proposed approaches.
Such as Least Confident ~\citep{culotta2005reducing} (Eq~\ref{tab:leastconfident}), 
\begin{equation}
\phi^{LC}(\mathbf{x}) = 1 - P(y^* \mid \mathbf{x}; \theta).
\label{tab:leastconfident}
\end{equation}
Margin Sampling~\citep{scheffer2001active} (Eq~\ref{tab:marginsampling}), 
\begin{equation}
\phi^{M}(\mathbf{x}) = -\left(P(y_1^* \mid \mathbf{x}; \theta) - P(y_2^* \mid \mathbf{x}; \theta)\right).
\label{tab:marginsampling}
\end{equation}
and Max Entropy ~\citep{kim2006mmr} (Eq~\ref{tab:entropy})
, 
\begin{equation}
\phi^{ME}(\mathbf{x}) = -\sum_{\hat{y} \in \mathcal{N}} P(\hat{y} \mid \mathbf{x}; \theta) \log P(\hat{y} \mid \mathbf{x}; \theta).
\label{tab:entropy} 
\end{equation}

\textit{LSB}: 
The 130k LSB data is obtained by retrieving the top $k=3$ nearest neighbors of a set of 50k seed badcases from 60 million unlabeled candidates based on their CLS hidden embeddings. 
A subset of 30k is randomly sampled from 130k LSB data for later experiment.


\textit{LSB w/ LT}: 
The 30k LSB-LT data is selected from the 130k LSB dataset by applying Lookalike threshold (LT) $0.5$.


\textbf{Clickbait}:
Choosing from the past 60 days video logs, 
the dataset includes 550k randomly sampled data, 260k statistical AL and 40k kNN-LSB.

For comparison, we pre-train and warmed-up the model by 500k random samples as a fixed snapshot, and the remaining 50k samples for each group are selected using different active learning (AL) methods to evaluate the effectiveness of various strategies:
\textit{Random}: the remaining 50k from the randomly sampled set.
\textit{Statistical AL}: 50k sampled from 260k Stat AL set.
\textit{LSB}: 10k sampled from 40k LSB set.
\textit{LSB w/ LT}: 10k selected from 40k LSB by applying LT.

\textbf{ANSA}:
Choosing from the past 90 days video logs, 
the dataset includes 800K randomly sampled data, 400K statistical AL data, and 60K kNN-LSB data. For comparison: 
\textit{Random}: 300k randomly sampled from 800k.
\textit{Statistical AL}: 300k (sampled from 400k Stat AL set).
\textit{LSB}: 50k (sampled from 60k LSB set).
\textit{LSB w/ LT}: 50k (selected from 60k LSB by apply LT).




\subsubsection{Audio}: For all above clickbait training data and the same evaluation data, an audio feature is added as an extra domain. 
Raw audio information uses WAV format, processed into 80*3000 (audio's frequency dimension and time step) 2D Mel-spectrogram vectors with 60s duration and 16,000 Hz sample-rate,  as input for Whisper-small encoder.

Based on the original 40k randomly sampled data, add 20k audio-sensitive samples from the past 60 days to form another evaluation set to test the impact of VLMAE.

\subsection{Model Training Details}
The multimodal search relevance model (SRM) or feed quality models (ANSA/ Clickbait etc.) usually employ a BEiT3~\cite{wang2022image} or VLMo-like ~\cite{li2021vlmo} architecture, which adopts a mixture-of-modality-experts transformer with bidirectional attention to handle multimodality input, i.e., search query, text information(title, sticker), and video frames.

\textbf{Search Relevance Model.}
The backbone employs a 12-layer VLMo-Albert, combined with a 3-layer 4-class MLP trained on 4 NVIDIA A100-80GB GPU. 
The model integrates features from several textual domains, including search query, title, category, etc., and video patch embedding domain. Each text domain is tokenized with a maximum length of 80. 
The model is optimized by an Adam optimizer with a fixed learning rate of 4e-5 and a batch size of 64. Training takes up to 5 epochs and can be terminated earlier if the test accuracy stabilizes. 

\textbf{Clickbait Model.} Employs 24/48-layer VLMo-Albert model defined as \textit{small/large}, combined with a 3-layer 4-class MLP trained on 32 NVIDIA A100-80GB GPU. Features include text title, sticker, and video patch embedding. Optimizer and other training settings are the same as the Search Relevance's.

\textbf{ANSA Model.} Employs a 48-layer VLMo-Albert model, combined with a 3-layer binary classification MLP trained on 16 NVIDIA A100-80GB GPU. Other settings are the same as the Clickbait's.

\textbf{Lookalike Model.} 
The binary classification model for the lookalike threshold is trained separately on the training sets of the above models, using a logistic regression model (LR). 
Based on the description in Sec~\ref{sec:lookalike}, positive samples are defined as seed badcases, and negative samples are selected from those where model predictions match the labels. 
The search-ads LT model achieves an AUC score of 0.7031, while the ANSA LT model's AUC score is 0.8805, and the clickbait LT model's AUC score is 0.6347.

\textbf{Audio Encoder.} The audio encoder  employs Whisper-Small\cite{radford2022whisper} 's 12-layer encoder with 768 hidden dimensions, which is a transformer-based architecture with 244M parameters designed for multilingual speech-to-text and translation tasks.


\subsection{Metrics}

\begin{table}
  \centering
      \scriptsize 
  \setlength{\tabcolsep}{7pt} 
  \resizebox{\linewidth}{!}{
    \begin{tabular}{l|ccc}
    \hline
    \hline
    \textbf{Strategy} & \textbf{Search Ads} & \textbf{Clickbait} & \textbf{ANSA} \\
    \hline
    Random &0.083& 0.069 & 0.068  \\
    LeastConfidence &0.303& 0.259 & 0.317  \\
    MarginSampling &0.258& 0.297 &- \\
    MaxEntropy &0.297& 0.345 &-  \\
    \textbf{LSB-kNN} & \textbf{0.380} & \textbf{0.346} & \textbf{0.329}   \\
    \textbf{LSB-kNN-LT} & \textbf{0.400} & \textbf{0.409} & \textbf{0.421}  \\
    \hline
    \hline
    \end{tabular}
  }
  \caption{Comparison of various strategies based on multi-class loss for the Search Ads (4-class), Clickbait (4-class), and ANSA (binary-classification) tasks. LSB-kNN selected cases have higher multi-class loss, suggest the data selected are more difficult for model to learn. LT further enhances the quality of kNN-selected cases with search ads > 0.5, clickbait > 0.3, and ANSA > 0.3.}
  \label{tab:loss-compare}
\end{table}

\subsubsection{Effectiveness of Dataset}
We compare the difference between human annotation and model prediction as $multi-class\; loss$ on different datasets listed above. 

\textbf{Multi-class Loss} corresponds to the average difference between the expectation of the prediction probability of the model and the human label. 
For a dataset that includes $k$ data, each data's multi-classification labels ranging from 0-$n$. 
The higher loss means that the dataset is more potentially valuable for further model training. (Eq \ref{loss}) 


\begin{equation}
\mathrm{Loss} = \frac{1}{k} \sum_{j=1}^{k} \left| \mathcal{M}[X_j] - y_j \right|
\label{loss}
\end{equation}

\begin{table*}[ht]
\centering
    \scriptsize 
  \setlength{\tabcolsep}{5pt} 
\resizebox{\textwidth}{!}{%
\begin{tabular}{c|ccccccccc}
\hline
\hline
\textbf{Category} & \textbf{\shortstack{Strategy}} & \textbf{\shortstack{AUC}} & \textbf{\shortstack{F1-score}}  &   \textbf{\shortstack{R@P40}}&  \textbf{\shortstack{R@P50}} & \textbf{\shortstack{R@P60}} & \textbf{\shortstack{R@P70}} & \textbf{\shortstack{R@P80}} & \textbf{\shortstack{R@P90}} \\

\hline
Search Ads & Random 160k & 0.969 & 0.957 
&0.872 & 0.842 & 0.811 & 0.727 & 0.498 & 0.261 \\
 & Stat AL 160k & 0.973 & 0.958 &0.889& 0.851 & 0.786 & 0.746 & 0.521 & 0.284 \\
 & \textbf{Stat AL 130k + KNN 30k} & \textbf{0.976} & \textbf{0.964} &\textbf{0.903}& \textbf{0.878} & \textbf{0.832} & 0.763 & 0.59 & 0.342 \\

 & \textbf{Stat AL 130k + KNN 30k(LT)} & 0.974 & 0.962 &0.887& 0.868 & 0.828 & \textbf{0.769} & \textbf{0.616} & \textbf{0.349} \\
\hline
Clickbait & Random 50k & 0.851 & 0.879  &0.721& 0.65 & 0.577 & 0.481 & 0.331 & 0.0486 \\
 & Stat AL 50k & 0.857 & 0.886 &0.733& 0.668 & 0.585 & 0.503 & 0.356 & 0.102 \\
 & \textbf{Stat AL 40k + KNN 10k} & \textbf{0.857} & 0.887 &\textbf{0.745}& 0.674 & 0.598 & \textbf{0.519} & 0.372 & 0.108 \\
 & \textbf{Stat AL 40k + KNN 10k(LT)} & 0.856 & \textbf{0.887} &0.742& \textbf{0.675} & \textbf{0.603} & 0.507 & \textbf{0.377} & \textbf{0.118} \\
\hline
ANSA & Random 300k & 0.906 & 0.779  &0.928& 0.902 & 0.844 & 0.794 & 0.744 & 0.685 \\
 & Stat AL 300k & 0.920 & 0.795 &0.954& 0.912 & 0.866 & 0.824 & 0.776 & 0.696 \\
 & \textbf{Stat AL 250k + KNN 50k} & 0.918 & 0.798 & \textbf{0.961} & \textbf{0.914} & 0.863 & 0.820 & 0.779 &0.716 \\
 & \textbf{Stat AL 250k + KNN 50k(LT)} & \textbf{0.923} & \textbf{0.801} &0.9611& 0.914 & \textbf{0.869} & \textbf{0.825} & \textbf{0.779} & \textbf{0.721} \\
\hline
\hline
\end{tabular}%
}
\caption{
Comparison of various data selection strategies for comparison. The strategies include random sampling, statistical active learning, LSB-kNN, and LSB-kNN after LT selection. The results show that models trained with KNN-filtered data using a LT outperformed the others across all categories: search ads, clickbait, and ANSA. For these issues, since the online threshold requires high-accuracy recall to prevent false positives, the improvement of recall under high precision by LT is highly significant. The maximum BETA variances for clickbait and ANSA are under $10^{-4}$ at 90\% precision, while for search ads is under $10^{-3}$ at 90\% precision. 
}
\label{tab:strategy-comparison}
\end{table*}

\begin{table*}[h!]
  \centering
  \resizebox{\textwidth}{!}{%
  \begin{tabular}{l|cccl}
    \hline
\hline
    \textbf{Search Query} & \textbf{Label}  & \textbf{Four-class prediction} & \textbf{Pair Ad's Industry} & \textbf{Explaination} \\
    \hline
    \textbf{oxygen} & 0  & 0.164, 0.003, 0.110, 0.721 & Others & Oxygen - Health Products \\
    Motorcycle & 0 & 0.162, 0.002, 0.125, 0.709 & Other Automotive Aftermarket & Motorcycle - Helmet \\
    glass cleaner & 2 & 0.158, 0.003, 0.095, 0.741 & Car Cleaning & Glass Cleaner \\
    activities for 2 year toddler & 0  & 0.190, 0.002, 0.119, 0.686 & Video Products & 2-Year-Old Mobile Games \\
    \hline
    \textbf{mukbangs} & 3  & 0.495, 0.004, 0.156, 0.343 & Food and Beverage E-Commerce & Mukbang - Health Products/Snacks \\
    easy glue on nails & 3  & 0.489, 0.004, 0.171, 0.334 & Nail Care & Easy Glue on Nails \\
    mystery boxes & 3  & 0.478, 0.007, 0.148, 0.372 & Comprehensive 2C E-commerce & Mystery Boxes \\
    people with funny laughs & 3  & 0.508, 0.004, 0.152, 0.335 & Short Video & Funny Video \\
    \hline
    \textbf{michael scott costume} & 2 & 0.813, 0.022, 0.090, 0.073 & Women's Clothing & Comedy Cosplay - Fashion \\
    Diy craft table & 1  & 0.812, 0.022, 0.092, 0.064 & Furniture & DIY - Lifting Table \\
    NICHOLAS CHAVEZ COSTUME & 2  & 0.803, 0.029, 0.101, 0.067 & Clothing Accessories E-Commerce & Comedy Cosplay - Fashion \\
    bumper cars for kids & 2  & 0.807, 0.025, 0.085, 0.081 & Toys & Bumper Cars \\
    \hline
\hline
  \end{tabular}}
    \caption{\label{citation-guide}
    LSB case study for Search Ads: each cell's first row is seed badcase, and the following three cases are expanded by LSB. 
    It is possible to find semantically similar cases as "michael scott costume" to "NICHOLAS CHAVEZ COSTUME". 
    Similarly, in cases of underestimation, the multiclass probability distribution assigns a probability of 0.8 to a score of $0$ ("completely irrelevant"), while manual annotations consistently assign a score of $2$ ("moderately relevant"). 
    This also applies to the underestimation in the second group and the overestimation in the first group.
  }
  \label{tab:kNN_case}
\end{table*}

\subsubsection{Offline Metrics of Model}
For clickbait problem, "0,1,2,3" labeled as "normal", "borderline", "non-detective", "detective". 
For ANSA problem, "0,1" stands for "not ANSA" or "ANSA", 
while for search ads, "0,1,2,3" refers to query-ads "completely irrelevant," "slightly relevant,"  "moderately relevant," and "highly relevant".
We utilize following metrics to evaluate the offline performance of the model in the evaluation dataset:


\textbf{AUC, F1-score, Recall at various Precision thresholds} are standard metrics for binary classification evaluation. 
For all three tasks, we base on zero or non-zero scores to calculate F1, AUC, and precision-recall rate.

\textbf{Beta Variance} is defined as the largest variance of a Beta distribution used to model the posterior uncertainty of the positive-class proportion. \cite{NEURIPS2018_a981f2b7}
Concretely, if a subset of samples has \(\alpha\) for true positives amount + 1 and \(\beta\) for false positives amount + 1, we consider the Beta posterior as
\begin{equation}
  \text{Variance} = \frac{\alpha \beta}{(\alpha + \beta)^2 (\alpha + \beta + 1)}
\end{equation}
Then compute this variance for each precision/recall subset and take the maximum as the \emph{max Beta Variance}, reflecting the highest level of uncertainty in estimating the statistic significance of the precision-recall value.
\emph{We consider that maximum Beta Variance below 0.5\% suggests strong certainty.}

\subsubsection{Online Metrics} \

\textbf{Search Ads.}  Evaluated by metrics as platform advertising delivery (Total Ad Impressions) and revenue (Total Ad Deductions).

\textbf{ANSA.} For post-launch performance, we primary focus on online metrics like inappropriate content view rate (ICVR), and sexual suggestive view rate (SSVR).

\section{Results}

\subsection{LSB Performance}

We evaluate the LSB strategy against statistically-based active learning from three perspectives: (1) the quality of selected samples, (2) impact of selected samples on model training performance, and (3) case analysis.

\subsubsection{Quality of Sample Selection}
In Table \ref{tab:loss-compare}, through the multiclass loss analysis, kNN exhibits significantly higher overall loss compared to statistical active learning (AL) strategies in search ads, Clickbait and ANSA tasks. 
This can be attributed to the selection of seed bad cases based on high-loss thresholds, with kNN identifying candidates that share similar CLS embeddings, which are often associated with prediction errors. 
Furthermore, kNN demonstrates a stronger capability in capturing samples with high prediction confidence but incorrect output (e.g., [0.1, 0.1, 0.1, 0.7]), resulting in a larger pool of candidate samples compared to those selected by statistical AL methods. 
Further filtering of kNN candidates using the Lookalike threshold (LT) leads to an additional increase in the quality of selected samples, indicating that LT effectively refines kNN-selected candidates into distinct badcase patterns.

\begin{table*}[th!]
\centering
\resizebox{\textwidth}{!}{%
\begin{tabular}{c|cccccccc|cccccc}
\hline
\hline
\multicolumn{1}{c|}{} & \multicolumn{8}{c|}{\textbf{Clickbait Test-set}} & \multicolumn{6}{c}{\textbf{Clickbait More Audio Test-set}} \\ 
\cline{2-15}
 \textbf{Model} & \textbf{AUC} & \textbf{F1}  & \textbf{R@30} & \textbf{R@40} & \textbf{R@50} & \textbf{R@60} & \textbf{R@70} & \textbf{R@80} & \textbf{AUC} & \textbf{F1}  & \textbf{R@60} & \textbf{R@70} & \textbf{R@80} & \textbf{R@90} \\

\hline
  VLMo-small & 0.925 & 0.622  & 0.792 & 0.737 & 0.696 & 0.626 & 0.549 & 0.412 & 0.643 & 0.458  & 0.523 & 0.431 & 0.381 & 0.335  \\
  VLMo-small avg-pool &  0.925 & 0.624 & 0.779 & 0.729 & 0.678 & 0.630 & 0.562 & 0.437 &  0.689 & 0.450  & 0.607 & 0.408 & 0.371 & 0.337 \\
  \textbf{VLMAE-small} & \textbf{0.925} & \textbf{0.645}  & \textbf{0.790} & \textbf{0.737} & \textbf{0.692} & \textbf{0.655} & \textbf{0.602} & \textbf{0.505} & \textbf{0.717} & \textbf{0.463}  & \textbf{0.713} & \textbf{0.455} & \textbf{0.397} & \textbf{0.357}  \\
\hline
  VLMo-large  & 0.945 & 0.682  & 0.851 & 0.811 & 0.771 & 0.732 & 0.672 & 0.54 & 0.747 & 0.463  & 0.8 & 0.497 & 0.428 & 0.38  \\
  VLMo-large avg-pool & 0.940 & 0.665  & 0.844 & 0.807 & \textbf{0.779} & \textbf{0.733} & 0.664 & 0.523 & 0.735 & 0.494  & 0.755 & 0.554 & 0.466 & 0.403 \\
  \textbf{VLMAE-large} & \textbf{0.944} & \textbf{0.688}  & \textbf{0.853} & \textbf{0.818} & 0.777 & 0.73 & \textbf{0.683} & \textbf{0.577} & \textbf{0.773} & \textbf{0.549}  & \textbf{0.888} & \textbf{0.582} & \textbf{0.484} & \textbf{0.435}  \\
\hline
\hline
\end{tabular}%
}
\caption{Performance comparison across models on the Clickbait VLMo model using 40k random samples, with an additional 20k Audio-Sensitive test samples. Small refers to the 24-layer VLMo, while Large refers to the 48-layer VLMo. The maximum BETA variance on the standard testset is under $10^{-4}$, whereas for the extended audio-sensitive testset's under $10^{-5}$.}
\label{tab:ansa_clickbait_audio}
\end{table*}

\subsubsection{Model Performance}
As demonstrated in Table \ref{tab:strategy-comparison}, with the same amount of training data, the models trained in the samples selected by all active learning (AL) strategies significantly outperformed those trained on randomly selected samples. 

In the third row of comparison, by replacing a portion of the statistical AL data with the LSB-selected data, both the recall rate and the F1 score show improvement. This indicates that a small amount of LSB data can partially compensate for the distributional shortcomings of statistical AL by supplementing cases such as high confidence or seed badcase similar pattern samples.

In the fourth row of comparison, replacing randomly sampled LSB-kNN data with LSB data selected by the Lookalike model with threshold(search ads>0.5, clickbait>0.3, ANSA>0.3) leads to further metrics improvements, indicates that the filtered patterns more confidently identified as badcase samples provide greater value to the model.

Overall, when mixed with statistical AL, the LSB strategies can further improve model performance with same amount of training data. The Lookalike threshold(LT) can enhance the quality of selected data on top of the LSB strategy.

\subsubsection{Case Study}

According to Table \ref{tab:kNN_case}, in the search ads relevance task, when k=3 for kNN, among the 4,147 seed-badcase and 12,441 kNN-selected samples, we observe the following.

1. The score distribution of the bad cases selected by LSB-kNN is highly consistent with that of the seed bad cases, and the analysis shows that the badcase rate of the selected samples from LSB-kNN is 76\%. 
Since the seed badcases include high-confidence predictions, the kNN algorithm can also select similar samples, which would not be selected by statistical AL methods. 
For example, in the first group, the kNN-selected query-ad pair "Motorcycle - Helmet" had a prediction of 0.709 for "highly relevant", showing high confidence, but is actually incorrect. Its seed case, "Oxygen - Health Products," also has a predicted probability with same pattern.

2. LSB-kNN not only selects cases with a similar multi-class probability distribution but also captures semantically similar cases. For example, in the last group, the seed query-doc pair is "[michael scott costume - clothing]" with a predicted probability of 0.813 for "completely irrelevant", which is underestimated. The kNN algorithm selects the pair "[NICHOLAS CHAVEZ COSTUME - clothing]," which closely matches the semantic structure of the seed badcase which is also underestimated.

\subsection{VLMAE Performance}

We test audio feature fusion on the 24-layer and 48-layer VLMo models for the clickbait tasks using two methods: the genuine approach based on average pooling~\cite{dong2024coefvq} and VLMAE, evaluating their performance on both randomly sampled test sets and audio-sensitive test sets. 
Compared to models without audio features, the performance of the avg-pooling method is unstable across small/large models on two testsets. In contrast, the VLMAE group consistently outperformed all other groups. This indicates that the inclusion of audio features has the potential to enhance the model’s ability to handle audio-sensitive cases, and the VLMAE provides more significant gains when information from different modalities undergoes more effective feature fusion.

\subsection{Post-launch Performance}
For search ads, we conduct A/B experiments lasting 8 days with 5\% of total traffic, yielding a p-value of $1.467\%$. 
A model trained with 160k Stat-AL data achieved a 1.479\% increase in revenue compared to a model trained with an equal amount of randomly labeled data, while maintaining the same ad send volume. 
A model trained with an equal amount of LSB-LT data achieved a 2.638\% revenue increase under the same conditions. 
This indicates that, compared to conventional statistical active learning, the data collected and labeled using LSB-LT can significantly enhance the relevance model's performance, leading to the selection of more query-relevant ads and higher post-conversion rates.

For ANSA, we conduct A/B experiment lasting for 24 days with 15\% of total traffic, yielding a p-value of $0.1\%$.
The performance of the ANSA model improves after switching from baseline to LSB-LT + VLMAE, with online ICVR decreased by $0.226\%$ and SSVR decreased by $1.36\%$ 
, indicating that the upgraded model has a stronger capability to keep users from negative experience.


\section{Conclusion}
In modern industrial-scale deep neural network applications, traditional active learning methods based on statistical information entropy—such as least confidence, margin sampling, and entropy-based approaches—often suffer from the loss of semantic information and the oversight of high-confidence mispredictions. To address these limitations, the proposed kNN-based Latent Space Broadening method effectively expands the candidate pool for active learning, ensuring a more comprehensive selection of informative samples. Additionally, the integration of lookalike filtering and VLMAE enables a more robust and scalable approach for multimodal retrieval and recommendation scenarios. Empirical results from both offline and online experiments demonstrate that this method not only improves real-world deployment accuracy but also significantly enhances manual annotation efficiency, maintaining high model performance.


\bibliographystyle{ACM-Reference-Format}



\appendix

\begin{table*}[ht]
\centering
  \scriptsize 
  \setlength{\tabcolsep}{5pt} 
\resizebox{\textwidth}{!}{%
\begin{tabular}{c|c|cccccccccc}
\hline
\hline
\textbf{Category} & \textbf{Model} & \textbf{AUC} & \textbf{F1-score} & \textbf{R@10} & \textbf{R@20} & \textbf{R@30} & \textbf{R@40} & \textbf{R@50} & \textbf{R@60} & \textbf{R@70} & \textbf{R@80} \\
\hline
\multirow{7}{*}{ANSA} 
 & X-VLM\cite{wang2022multi} & 0.916 & 0.484 & 0.916 & 0.785 & 0.684 & 0.584 & 0.402 & 0.336 & 0.216 & 0.135 \\
 & VLMo\_8f\_12layer & 0.789 & 0.217 & 0.602 & 0.219 & 0.113 & 0.051 & 0.026 & 0.001 & 0.001 & 0.001 \\
 & VLMo\_8f\_24layer & 0.930 & 0.499 & 0.950 & 0.813 & 0.712 & 0.6108 & 0.470 & 0.327 & 0.244 & 0.059\\
 & VLMo\_8f\_48layer & 0.929 & 0.499 & 0.950 & 0.814 & 0.712 & 0.611 & 0.470 & 0.327 & 0.244 & 0.057\\
 & VLMo\_16f\_12layer & 0.781 & 0.204 & 0.590 & 0.199 & 0.097 & 0.037 & 0.001 & 0.001 & 0.001 & 0.001\\
 & VLMo\_16f\_24layer & 0.911 & 0.456 & 0.909 & 0.773 & 0.636 & 0.516 & 0.398 & 0.275 & 0.196 & 0.120\\
 & VLMo\_16f\_48layer & \textbf{0.931} & \textbf{0.524} & \textbf{0.946} & \textbf{0.829} & \textbf{0.737} & \textbf{0.648} & \textbf{0.547} & \textbf{0.401} & \textbf{0.291} & \textbf{0.175}\\
\hline
\multirow{7}{*}{Clickbait} 
 & Roberta\cite{liu2019roberta} & 0.917 & 0.614 & 0.945 & 0.855 & 0.776 & 0.716 & 0.669 & 0.617 & 0.519 & 0.417\\
 & VLMo\_8f\_12layer & 0.910 & 0.577 & 0.932 & 0.809 & 0.726 & 0.662 & 0.609 & 0.556 & 0.498 & 0.422\\
 & VLMo\_8f\_24layer & 0.925 & 0.622 & 0.944 & 0.856 & 0.792 & 0.737 & 0.696 & 0.626 & 0.549 & 0.412\\
 & VLMo\_8f\_48layer & 0.945 & 0.682 & 0.967 & 0.900 & 0.851 & 0.811 & 0.771 & 0.732 & 0.672 & 0.540\\
 & VLMo\_16f\_12layer & 0.910 & 0.578 & 0.931 & 0.819 & 0.735 & 0.680 & 0.625 & 0.569 & 0.507 & 0.402\\
 & VLMo\_16f\_24layer & 0.927 & 0.635 & 0.943 & 0.859 & 0.806 & 0.761 & 0.714 & 0.662 & 0.587 & 0.463\\
 & VLMo\_16f\_48layer & \textbf{0.945} & \textbf{0.689} & \textbf{0.968} & \textbf{0.903} & \textbf{0.861} & \textbf{0.821} & \textbf{0.788} & \textbf{0.749} & \textbf{0.678} & \textbf{0.580} \\
\hline
\hline
\end{tabular}%
}
\caption{
Performance comparison across models on the ANSA and Clickbait datasets. Metrics (AUC, F1-score, Recall rate at varying precision @10 - R@80) improved more by layers and parameter increacing comparing to frame increasing. 24-layer VLMo also usually better than Roberta or X-VLM.
}
\label{tab:ansa_clickbait_results}
\end{table*}

\section{More Ablation Analysis}

To eliminate the influence of video modality frame sampling and model layers, we conduct ablation experiments on the VlMo framework for the ANSA and clickbait tasks, comparing the impact of different video extract-frame numbers and model layers on the same dataset. The experiments demonstrated that, for both tasks, the impact of model layers is greater than the influence of video frame numbers. The offline performance with 16 frames is better than that with 8 frames (Table \ref{tab:ansa_clickbait_results}). Also, as the model layers and parameters increase, the improvement in matrix is more pronounced at different accuracy levels. The improvement on recall rate of about 3\%-5\% can be achieved through VLMAE multi-modality and LSB annotation optimization, while increasing the number of layers and parameters can lead to a 10\% improvement on recall rate. 
More than 24-layer VLMo also usually better than Roberta or BEIT-v2.

\end{document}